\documentclass[preprint,aps,prd,amsmath,showpacs,nofootinbib]{revtex4}
\usepackage{graphicx}

\newcommand{\beq}{\begin{equation}}
\newcommand{\eeq}{\end{equation}}
\newcommand{\bea}{\begin{eqnarray}}
\newcommand{\eea}{\end{eqnarray}}

\newcommand{\Ref}[1]{(\ref{#1})}

\begin{document}

\title{Stability of composite vacuum Heckmann wormholes in Brans-Dicke theory}

\author{Sergey M. Kozyrev}
\email{Sergey@tnpko.ru}
\affiliation{Scientific center for gravity wave studies ``Dulkyn'', Kazan, Russia}

\begin{abstract}
This paper discusses linearized (spherically symmetric)
perturbation of static Heckmann composite thin shell wormholes in
Brans-Dicke gravity. The equation of state $P= \beta^2 \sigma$ at
the throat is linearized around the static solution where $\sigma$
energy density of the shell and $P$ the lateral presume. We have
shown that this thin shell wormholes is stable within the range
$0\leq\beta<1$ and with all values of $\omega$ except $\omega=-2$.
\end{abstract}


\maketitle

\section{Introduction}
After the leading work by Moris and Thorn \cite{Thorn}, wormhole
connecting two asymptotically flat spacetimes is now well known
classic in relativity theories. A powerful theoretical method for
constructing and analyze wormhole geometry, introduced by Visser
\cite{Vis1,Vis2}, is by cutting and pasting two manifolds to form
a geodetsically complete one. It is well known that wormhole
within the framework of general relativity always requires the
presence of exotic matter \cite{Vis1}. On the other hand wormholes
in some alternative theories of gravity have been also
investigated previously and it was demonstrated that the
requriment of exotic matter can be avoided.

To describe a thin shell Brans-Dicke wormhole configurations one
may construct such system by two distinct strategies. Actually,
the essence of the standard method \footnote{The first examples of
thin-shell wormholes have been given by Visser \cite{Vis1,Vis2}.}
is the following: One takes two the same copies of spacetime
manifolds with appropriate asymptotics, cuts and casts away
'useless' regions of spacetimes (containing horizons,
singularities, etc.), and pastes remaining regions. As the result,
one obtains a geodesically complete wormhole spacetime with given
asymptotics (Schwarzschild, Reissner-Nordstrom, Brans-Dicke, etc.)
and a throat being a thin shell of exotic matter violating the
null energy condition. In other case, we have initially a thin
shell made from \textit{ordinary} matter, and then we look for
appropriate interior and exterior Brans-Dicke solutions \cite{BD,
Brans} matched at the shell \cite{Sushkov}.

Here it is worth noticing that thin-shell Brans-Dicke wormholes
were studied in the literature \cite{thin1,thin2}.
  Note
that this approach is similar to the problem of a thin shell in
general relativity (see Ref. \cite{thinshellingr}). However, the
distinction is that the Birkhoff theorem is not valid in
Brans-Dicke theory, and so both the interior and exterior
Brans-Dicke vacua are not unique.

It is then important to know wether such wormhole are stable.
Stability analysis of thin shell wormholes under small
perturbations around the static solution has been done previously
by several authors. In this paper, the stability of composite
vacuum Brans-Dicke wormholes is analyzed under linear perturbation
preserving the symmetry.

\section{Construction of the Brans-Dicke composite thin shell wormholes}
The action of Brans-Dicke theory \cite{FM} is given
by\footnote{Units $8\pi G=c=1$ are used throughout the paper.
Greek indices range over the coordinates of the 4-manifold and
Roman indices over the coordinates of the 3-surfaces.}
\beq\label{action} S =\frac12\int dx^4 \sqrt{-g}\left\{\phi R -
\omega\frac{\phi^{,\mu}\phi_{,\mu}}{\phi}\right\} +{S}_m, \eeq
where $R$ is the scalar curvature, $\phi$ is a scalar field,
$\omega$ is a dimensionless coupling parameter, and ${S}_m$ is an
action of ordinary matter (not including the scalar field). The
action \Ref{action} provides the following field equations:
\begin{subequations} \label{BDeqs}
\bea G_{\mu \nu } &=& \frac1\phi T_{\mu\nu}+ \frac{\omega}{\phi
^2}\phi_{,\mu}\phi_{,\nu} -\frac{\omega}{2\phi ^2} g_{\mu \nu
}\phi_{,\alpha}\phi^{,\alpha} + \frac 1\phi \phi_{;\mu;\nu} -\frac
1\phi g_{\mu \nu }\phi_{;\alpha}^{;\alpha},
\\
\phi^{;\alpha}_{;\alpha} &=&\frac{T}{2\omega+3}, \eea
\end{subequations}
where $G_{\mu \nu }=R_{\mu \nu }-\frac12 Rg_{\mu \nu }$ is the
Einstein tensor, and $T=T^\mu_\mu$ is the trace of the matter
energy momentum tensor $T_{\mu\nu}$.

We shall consider the matching of two static and spherically
symmetric spacetimes. From the mathematical point of view, the
most simple and satisfactory expression for the matching
conditions is, following Linchnerowicz, the assumption that there
exists a system of co-ordinates in which the metric tensor
satisfies the continuity conditions. Let $\xi^i $ be a coordinate
system on $\Sigma$ where $\Sigma$ is an abstract copy of any of
the boundaries. Continuity conditions  require a common coordinate
system on $\Sigma$ and this is easily done if one can set $\xi_+^i
= \xi_-^i$. 

Let $(V^{\pm},g^{\pm})$ be four-dimensional spacetimes with
non-null  $\Sigma^{\pm}$. The junction/shell formalism constructs
a new manifold $\cal M$ by joining one of the distinct parts of
$V^+$ to one of the distinct parts of $V^-$ by
the identification $\Sigma^+ = \Sigma^- \equiv \Sigma$. The matching conditions 
require the equality of the first and second fundamental forms on
$\Sigma^{\pm}$.
 There
are unique (up to orientation) unit normal vectors $
n_{\pm}{}^{\alpha}$ to the boundaries. We choose them so that if
$n_{+}{}^{\alpha}$ points towards $V^{+}$ then $n_{-}{}^{\alpha}$
points outside of $V^{-}$ or viceversa.  Clearly the sign of the
normal vectors are crucial since e.g. $n^-_\alpha$ points away
from the portion of $V^-$ which will be used in forming $\cal M$.
The three basis vectors tangent to $\Sigma$ are
\begin{equation}
e^\alpha_{i} = \frac{\partial x^\alpha}{\partial \xi^i}
\end{equation}
which give the induced metric on $\Sigma$ by
\begin{equation}
q_{i j} = \frac{\partial x^\alpha}{\partial \xi^i}
          \frac{\partial x^\beta} {\partial \xi^j} g_{\alpha \beta}.
\end{equation}
The extrinsic curvature (second fundamental form) is given by
\begin{eqnarray}
\label{eqn-Kij1} K_{i j} & = & \frac{\partial x^\alpha}{\partial
\xi^i}
              \frac{\partial x^\beta}{\partial \xi^j} \nabla_\alpha
n_\beta      \nonumber    \\
\label{eqn-Kij2} & = & -n_\gamma \left( \frac{\partial^2
x^\gamma}{\partial \xi^i \partial \xi^j} + {\Gamma^\gamma}_{\alpha
\beta}
  \frac{\partial x^\alpha}{\partial \xi^i}
  \frac{\partial x^\beta}{\partial \xi^j}
  \right).
\end{eqnarray}


Then matching conditions are simply
\begin{equation}
  q_{ij}{}^{+}=q_{ij}{}^{-},
\label{eq:backm}
\end{equation}
\begin{equation}
  K_{ij}{}^{+}=K_{ij}{}^{-}.
\label{eq:backmK}
\end{equation}

If both \Ref{eq:backm} and \Ref{eq:backmK} are satisfied we refer
to $\Sigma$ as a boundary surface. If only \Ref{eq:backm} is
satisfied then we refer to $\Sigma$ as a thin-shell.

 Writing two
static spherically symmetric spacetimes $V^+$ and $V^-$ with
signature $(- + + + )$. one can   suppose that the metrics
$g^+_{\alpha \beta}(x_+^\gamma)$ and $g^-_{\alpha
\beta}(x_-^\gamma)$ in the coordinate systems $x_+^\gamma$ and
$x_-^\gamma$ are of the most general forms
\begin{equation}
ds^2=-f^{\pm}(r,t)dt^2 + g^{\pm}(r,t) dr^2 + u^{\pm}(r,t)dr dt +
h^{\pm}(r,t) \left( d\theta^2+\sin ^2\theta d\varphi ^2 \right)
\label{eq:metricM}
\end{equation}
where $ f^{\pm}(r,t), g^{\pm}(r,t), u^{\pm}(r,t) $ and
$h^{\pm}(r,t)$ are of class $C^2$. However, in what follows, we
will restrict ourselves to static spacetimes with $u^{\pm}(r,t)=0
$ because that is sufficient for our illustrative purpose.

Within these spacetimes define two non-null 3-surfaces $\Sigma^+$
and $\Sigma^-$ with metrics $q^+_{ij}(\xi_+^k)$ and
$q^-_{ij}(\xi_-^k)$ in the coordinates $\xi_+^k$ and $\xi_-^k$
which decompose each of the 4-spacetimes into two distinct parts.
The parametric equation for hypersurface $\Sigma$ is of the form
\begin{equation}
\label{eqn-surf} F(x) = r-a =0. 
\end{equation}
in which case the throat is located at a constant radius $a$. The
norm to $\Sigma$ can be written as
\begin{equation}
n_\mu=-\sqrt{g(r)f(r)}\dot r dt|_{\Sigma}+\sqrt{g(r)(1+g(r)\dot
r^2)}dr|_{\Sigma} \,. \label{norm}
\end{equation}

The induced metric on the  $\Sigma$ by the two solutions
\Ref{eq:metricM} is
\begin{equation}
q^{\pm}_{i j} d \xi_{\pm}^i d \xi_{\pm}^j = -f^{\pm}(r)d\tau^2 +
h^{\pm}(r) \left( d\theta^2+\sin ^2\theta d\varphi ^2 \right)
\label{eq:metricIM}
\end{equation}
therefore we must have on the $\Sigma$ according to equality of
the first fundamental form
\begin{equation}
f^{-}(a)= f^{+}(a),   \quad  h^{-}(a)= h^{+}(a) \label{eq:tirstFF}
\end{equation}

 Note that this approach not required a continuity the
 $g^{\pm}(a)$. Moreover, Brans-Dicke scalar field should be
 continuous around the shell, however, field values on the inside
 and the outside of the shell do not necessarily have to be same
 \cite{FM}. We can impose that the $\Sigma$ be also
characterized by equality of the second fundamental form too. This
condition leads to
\begin{equation}
K_{1 1} = \frac{1}{2}\frac{f'(a)}{\sqrt{g(a)}}, \quad  K_{2 2} =
 \frac{1}{2}\frac{h'(a)}{\sqrt{g(a)}} \label{eq:tirstSF}
\end{equation}
where prime denote derivation wits respect to $a$. Continuity of
the second fundamental form  is merely equivalent to

\begin{eqnarray}
h^{+'}(a)^2 g^{-}(a) =h^{-'}(a)^2 g^{+}(a), \quad   f^{+'}(a)^2
g^{-}(a) = f^{-'}(a)^2 g^{+}(a)
 \label{eq:MC}
\end{eqnarray}

The junction conditions in Brans-Dicke theory (generalized
Darmois-Israel conditions) can be obtained by projecting on
$\Sigma$ the field equations \Ref{BDeqs} \cite{surfaceeqn}:
\beq\label{surfK}
-[K_j^i]+[K]\delta_j^i=\frac{1}{\phi}\left(S_j^i-\frac{S}{3+2\omega}\,\delta_j^i\right),
\eeq \beq\label{surfphi} [\phi_{,n}]=\frac{S}{3+2\omega}, \eeq
where the notation $[Z]=Z^{+}|_{\Sigma}-Z^{-}|_{\Sigma}$ stands
for the jump of a given quantity $Z$ across the hypersurface
$\Sigma$, $n$ labels the coordinate normal to this surface and
$S_{ij}$ is the energy-momentum tensor of matter on the shell
located at $\Sigma$. The quantities $K$ and $S$ are the traces of
$K^i_j$ and $S^i_j$ respectively. Note that Eq. \Ref{surfK} is
equivalent to \beq\label{surfS}
S^i_j=\phi\left(\frac{\omega+1}{\omega}[K]\delta^i_j-[K^i_j]\right).
\eeq

This Lanczos equations follow from the Einstein equation for the
thin shell.

\section{Stability condition (Linearizing method).}

We let the radius $a$ to be a function of the proper time $\tau$
on the $\Sigma$ and trying to find the local stability of the thin
shell under small perturbation around the static solution at
$a=a_0$. The jump of the components of the extrinsic curvature
associated with two sides of the hypersurface in the spacetime
with the metric \Ref{eq:metricM} can be found as

\beq K^\tau_\tau=\frac{  f(2\ddot{a} g+\dot{a}^2
g')+f'(1+\dot{a}^2 g) \big)   } {2 f\sqrt {g}\sqrt{1+\dot a^2 g} }
,\label{Ktau} \eeq

\beq
K_{\theta}^{\theta} = K_{\phi}^\phi = \frac{h'\sqrt{g(1+\dot{a}^2 g)}}{2gh}  \label{Ktheta}  \,, \\
\eeq

\beq K=K^\tau_\tau+2K^\theta_\theta. \eeq where the overdot
denotes a derivative with respect to $\tau$.

The surface stress-energy tensor of a perfect fluid is given by
\begin{equation} \label{Sperfluid}
S^i_j= \left(
\begin{array}{ccc}
-\sigma & 0 & \phantom{.}0 \\
0 & p & \phantom{.}0\\
0 & 0 & \phantom{.}p
\end{array}
\right),
\end{equation}
where $\sigma$ and $p$ are the surface energy density and the
surface pressure, respectively. So $S=2p-\sigma$. After some
algebra Eq. \Ref{surfS} yields



\bea \left[K_\tau^\tau\right]=
\frac{1}{(3+2\omega)\phi}\left[(2+\omega)\sigma+(2+2\omega)p\right]
\, \label{ik1}
 \eea \bea \left[K_\theta^\theta\right]=
-\frac{1}{(3+2\omega)\phi}(p+\sigma+\omega\sigma) \,.\label{ik2}
\eea

Rearranging the Eq.(\ref{ik2}) one
 can write

\begin{equation}
{1\over2} \dot{a}^2+V(a)=0  \,, \label{potencial}
\end{equation}

 By solving \Ref{ik2} for $\dot a^2$, one finds
\bea \dot a^2= \left(
\frac{2}{(3+2\omega)\phi}(p+\sigma+\omega\sigma)\frac{
h}{[h']}\right)^2-\frac{1}{g} \equiv -2 V(a) \,. \label{rdv} \eea

A Taylor expansion to second order to potential $V(a)$ around the
static solution yields:

 \beq V(a)=V(a_0) + V'(a_0)(a-a_0) + \frac{1}{2}V"(a_0)(a-a_0)^2 +
 O(a-a_0)^3.
\eeq


Differentiating (\ref{potencial}) with respect to $\tau$, we have
\beq V'(a)=-\ddot a \,. \eeq Solving (\ref{ik1}) for $\ddot a$, we
obtain
\beq V'(a_0)=\frac{\sqrt{1+\dot a^2 g}}{\sqrt{g}(3+2\omega)\phi}
\left(2P(1+\omega)+\sigma(2+\omega)\right)-\frac{ \dot a^2}{2}
\left(\frac{[f']}{f}+\frac{[g']}{g}\right) - \frac{[f']}{2 f g}.
\label{vpr} \eeq


It is clear that stress energy structure of the thin shell and its
property are related so that an imposed equation of state will
determine the stability condition. Because of its simplicity the
linear equation of state $p=\beta_0^2 \sigma$ is often used with
the field equations for many of applications  and it is frequently
used in constructing thin shell. Using the parameter $\beta_0 =
\frac{dp}{d\sigma}$ which is usually interpreted as a speed of
sound Eq. (\ref{surfphi}) and Eq. (\ref{vpr}) can be written as


\beq
 V'(a_0)= - \frac{\left(2+\omega +2 \beta ^2 (1+\omega
)\right)}{\left(-1+2 \beta ^2\right)}\frac{[\phi '] }{ \phi  g}
-\frac{ \dot a^2}{2} \left(\frac{[f']}{f}+\frac{[g']}{g}\right) -
\frac{[f']}{2 f g}  \label{vfirst} \eeq


Using Eq. (\ref{surfphi}) and Eq. (\ref{vpr}) it is not difficult
to see that $V(a_0) = V'(a_0) = 0$, so the potential
 \beq V(a)= \frac{1}{2}V"(a_0)(a-a_0)^2 + O(a-a_0)^3.
\eeq  Finally, by differentiating \Ref{vfirst} we find

\begin{eqnarray}
 V''(a_0) = \frac{\dot{a}^2}{2}\left(\frac{[f']^2}{f^2}-\frac{[f"]}{f}+\frac{[g']^2}{g^2}-\frac{[g"]}{g}\right) + \nonumber \\ + \frac{(2 + \omega +
 2 \beta^2 (1 + \omega)) }{g (2 \beta^2 - 1  )}\left(\frac{[\phi '] [g']}{\phi g}+\frac{[\phi ']^2}{\phi^2}-\frac{[\phi'']}{\phi}\right)
+\frac{1}{2g}\left(\frac{[f ']^2}{f^2}+\frac{[f'] [g']}{f
g}-\frac{[f'']}{f}\right) \label{vsecond}
\end{eqnarray}

The wormhole is stable if and only if $ V''(a_0)>0$.

\section{Spherically symmetric wormholes.}
  The matter is concentrated at the
throat, while the rest of the spacetime is the vacuum Brans-Dicke
solution can be given by


%
\begin{eqnarray}\label{BDmetricI}
 ds^2=-\nu_0 \left(1- \frac{2\mu}{\rho(r)}\right)
^{\frac 1 \lambda } dt^2 +\rho(r)'^2 \left(1-
\frac{2\mu}{\rho(r)}\right)^{-\frac {1+C} {\lambda} } dr^2 +
\nonumber
\\
+ \rho(r)^2 \left(1- \frac{2\mu}{\rho(r)}\right)^{1-\frac {1+C} {\lambda} }d\Omega^2, \\
  \phi(r) = \phi _0\left(1-
\frac{2\mu}{\rho(r)}\right) ^{\frac {C}{2\lambda} },
\end{eqnarray}
where $d\Omega^2=d\theta^2+\sin^2\theta d\varphi^2$ is the linear
element of the unit sphere, and $\rho(r)$ the arbitrary function.
Generally, this solution depends on free parameters: $\phi_0$,
$\nu_0$, $\mu$, and $C$. The parameter $\lambda$ is not free, it
obeys the following constraint condition: \beq \label{A}
\lambda=\sqrt{(C+1)^2-C\left(1-\textstyle\frac12\omega C\right)}.
\eeq

In this solution the interpretation to a choice of function
$\rho(r)$ generally correspond to a choice of state within the
"interior" and "exterior" vacua. In particular, considering the
function $\rho(r)$ in the form \beq \rho (r)=r, C =
\frac{2}{\omega}\eeq considerable simplification occurs, namely
the interior is described by the vacuum Heckmann solution
\cite{Heckmann} :
\begin{subequations} \label{interior}
\begin{eqnarray}
f^+(r) &=& \alpha_{0} \left(1- \frac{B}{r}\right) ^{\frac{\omega}{2+\omega} }, \\
g^+(r) &=& \frac{1}{1-\frac{B}{r}}, \label{glambda}\\
h^+(r) &=& r^2  \\
 \phi^+(r) &=& \phi_{0}\left(1- \frac{B}{r}\right) ^{\frac{1}{2+\omega} },
\end{eqnarray}
\end{subequations}
where $\phi_{0}$, $\alpha_{0}$,  $B$, and $C$ are free (still
undefined) parameters. Note that the radial coordinate $r$ runs
monotonically from $B$ to $a$, where $a>B$ is a boundary of the
interior region.

Assuming $C^{-} \equiv 0$ for exterior region of a spherical
gravitating configuration one can obtain the exterior
Schwarzschild solution:
\begin{subequations} \label{exterior}
\begin{eqnarray}
 f^-(r) &=&
1-\frac{2M}{r}\\
g^-(r)  &=&  \frac{1}{1-\frac{2M}{r}} , \\
h^-(r) &=& r^2  \\
\phi^-(r) &=& 1.
\end{eqnarray}
\end{subequations}
We consider a thin shell between two Heckmann manifolds with
exterior parameter $M$ (Schwarzschild mass) and interior mass
parameter $B$. The radial coordinate $r$ within the exterior
region runs from $a$ to infinity. We will suppose that $a>M/2$;
this guarantees that the exterior region does not contain the
event horizon.

So, $r$ is the global radial coordinate monotonically running from
$B$ to $a$ in the interior region, and from $a$ to infinity in the
exterior one. We will suppose that the all metric components and
scalar field be continuous on the $\Sigma$.



$ f^+(a)=f^-(a), \quad g^+(a)=g^-(a), \quad h^+(a) = h^-(a)$ and
$\phi^+(a)=\phi^-(a)\equiv 1. $

Obviously, for this case Darmois-Israel conditions \cite{DI}
holds. Substituting Eqs. \Ref{interior} and \Ref{exterior} gives
\begin{subequations}\label{coefs}
\bea
B &=& M,  \\
\alpha_0&=&\left(1-\frac{2M}{a}\right)^\frac{2}{2+\omega},\\
\phi_0&=&\left(1-\frac{2M}{a}\right)^{-\frac{1}{2+\omega}}. \eea
\end{subequations}
At the same time, derivatives of the metric and scalar field can
be discontinuous. The discontinuity of the metric is usually
described in terms of a jump of the extrinsic curvature $K_{ij}$.

    Finally, to find stable configurations, we need to compute $V''(a)$ from
 Eq. \Ref{vsecond}. Substituting Eq. (\ref{interior}) and Eq. (\ref{exterior}) we obtain

\beq V''(a_0)= 4 \mu \frac{ \mu \left(5+2 \beta ^2 (\omega-2
)+\omega \right)-a_0 \left(3+\omega +2 \beta ^2 \omega
\right)}{a_0^3 (2 \mu-a_0)  \left(2 \beta ^2-1\right) (2+\omega )}
\eeq


\section{Stability regions}
To find stable configurations, we need to consider all possible
constraints. The wormhole is stable if and only if \bea V''(a_0) >
0. \label{stable} \eea

The boundary of the region of stability is given by the surface
\bea V''(a_0, \omega, \mu) = 0. \label{stableS} \eea

We shell now study this equation in detail.  Under standard
physical circumstance the shell made from a perfect fluid with the
barotropic equation of state $p=\beta^2\sigma$ with $0\le
\beta<1$.
 To avoid singular behavior of the metric, the radius of the throat
must satisfy
\bea a_0 > 2 \mu. \label{abt} \eea

This conditions gives some possible cases accordingly to the value
of coupling constant $\omega$

1. Case $\omega<-2$

a. $2 \mu < a_0 \leq 3\mu$, and $a_0 > 3 \mu $,  $\omega \leq
\frac{3a_0-5 \mu}{\mu-a_0}$
    \bea \beta > \frac{1}{\sqrt{2}}. \label{abN} \eea

b. $a_0> 3 \mu $, $\frac{3 a_0-5 \mu }{\mu - a_0} < \omega < -2$
 \bea 0<\beta <\Upsilon \label{abB } \eea where $\Upsilon = \sqrt{\frac{a_0 (3+\omega )-\mu (5+\omega )}{2 \mu
(\omega -2)-2 a_0 \omega }}$

2.  Case $-2<\omega<-\frac{3}{2}$ \bea \Upsilon<\beta
<\frac{1}{\sqrt{2}} \label{abM} \eea

3.  Case $\omega=-\frac{3}{2}$ \bea \beta \neq \frac{1}{\sqrt{2}}
\label{abM} \eea

 4.  Case
$\omega>-\frac{3}{2}$

a. $ a_0 > \frac{7 \mu}{3}$, $  -\frac{3}{2} < \omega < \frac{2
\mu }{\mu - a_0}$
  \bea \frac{1}{\sqrt{2}}<\beta < \Upsilon \label{abM} \eea

b. $ a_0 > \frac{7 \mu}{3}$, $\omega \geq \frac{2 \mu }{\mu -
a_0}$
  \bea \beta >\frac{1}{\sqrt{2}} \label{abM} \eea



The interior and exterior regions are matched at a thin shell made
from ordinary matter with positive energy density and pressure.
The resulting configuration represents a stable composite
wormhole, i.e. the thin matter shell with the Schwarzschild-like
exterior region and the interior region containing the wormhole
throat.

\section{Summary}
In this paper, we have analyzed the stability of a new class of
static spherically symmetric configurations composed of interior
and exterior Brans-Dicke vacua divided by thin matter shell. Both
vacua correspond to the same Brans-Dicke coupling parameter
$\omega$, however they are described by the solution with
different sets of parameters of integration and arbitrary function
$\rho(r)$. In particular, the exterior vacuum solution has $C^-
\equiv 0$. In this case the solution with any $\omega$ just
reduces to the Schwarzschild one being consistent with
restrictions on the post-Newtonian parameters following from
recent Cassini data. The interior region possesses a strong
gravitational field, and so, generally, $C^{+}\not=0$.

The common geometry emerges from the junction conditions at the
boundary surface. It must be combined with Darmois-Israel junction
conditions. Our approach provides a clear way of showing that the
interior and exterior solutions is not a unique static spherically
symmetric solution.

An interesting feature of composite wormholes is that the
strong-field interior region containing all exotic ghost-like
matter is hidden behind the matching surface, whereas the
weak-field region out of it possesses the usual Schwarzschild
vacuum. Such the configuration is similar to the model of
trapped-ghost wormholes \cite{BroSus}. Note that in both models
wormholes are twice asymptotically flat. However, in the
trapped-ghost wormhole model the ghost is hidden in some
restricted region around the throat, whereas in the composite
wormhole model the ghost-like Brans-Dicke scalar occupies the
``half'' of wormhole spacetime behind the matching surface
\cite{Sushkov}. Anyway, in the composite wormhole configuration a
"ghost" is hidden in the strong-field interior region, which may
in principle explain why no ghosts are observed under usual
conditions.

Applying the analysis to composite wormhole geometries,
considering the one of the Heckmann solutions as interior region,
we deduced stability regions, and found that the latter exist well
within the wide range of parameters $0 \leq \beta<1$ and $\omega
\neq -2$. We would like to remark, that the Brans-Dicke thin shell
wormholes has been mentioned in \cite{thin2} with a stability
analysis. Actually, unlike authors claim in \cite{thin2}, a
stability region of Heckmann composite wormholes exist with all
values of $\omega$ except $\omega=-2$.

In this context the idea of linkages for vacuum regions bounding
other metrics may have broad applications.


\begin{thebibliography}{99}


\bibitem{BD}
C. Brans and R. H. Dicke, \textit{Mach's Principle and a
Relativistic Theory of Gravitation}, Phys. Rev. {\bf 124}, 925
(1961).

\bibitem{Brans}
C. H. Brans, \textit{Mach's Principle and a Relativistic Theory of
Gravitation. II} Phys. Rev. {\bf 125}, 2194 (1962).

\bibitem{BroSus}
K. A. Bronnikov and S. V. Sushkov, \textit{Trapped ghosts: a new
class of wormholes}, Class. Quant. Grav. {\bf 27}, 095022 (2010).

\bibitem{thin1}
E. F. Eiroa, M. G. Richarte, C. Simeone, \textit{Thin-shell
wormholes in Brans–Dicke gravity}, Phys.Lett. A373 (2008) 1-4;
Erratum-ibid. \textit{Erratum to "Thin-shell wormholes in
Brans–Dicke gravity"}, A373 (2009) 2399-2400.


\bibitem{FM}
Y. Fujii, N. Maeda, \textit{The scalar-tensor theory of
gravitation} (Cambridge, Cambridge University Press, 2003).

\bibitem{Heckmann}
O. Heckmann, P. Jordan, R.Fricke,  \textit{Zur erweiterten
Gravitationstheorie I.}, Z.Astroph., 28, 113-149, (1951).

\bibitem{thinshellingr}
A. P. Lightman, W. H. Press, R. H. Price, S. A. Teukolsky,
\textit{Problem book in relativity and gravitation} (Princeton,
Princeton University Press, 1975).

\bibitem{Thorn}
M.Morris, K. Thorne, U.Yurtsever,  \textit{Wormholes, Time
Machines, and the Weak Energy Condition. }Phys. Rev. Let. 61,
1446—1449 (1988).

\bibitem{DI}
N. Sen, Ann. Phys. (Leipzig) \textit{\"{U}ber die Grenzbedingungen
des Schwerefeldes an Unstetigkeitsfl\"{a}chen,} {\bf 73}, 365-396
(1924); K. Lanczos, \textit{Fl\"{a}chenhafte Verteilung der
Materie in der Einsteinschen Gravitationstheorie}, ibid. {\bf 74},
518–540 (1924); G. Darmois, \textit{M´emorial des Sciences
Math´ematiques}, Fascicule XXV, Chap. V (Gauthier-Villars, Paris,
1927); W. Israel, \textit{Singular hypersurfaces and thin shells
in general relativity,} Nuovo Cimento {\bf 44B}, 1, 1-14 (1966);
ibid. \textit{Singular hypersurfaces and thin shells in general
relativity,} {\bf 48B}, 2, 463 (1967).

\bibitem{surfaceeqn}
K.G. Suffern, \textit{Singular hypersurfaces in the Brans-Dicke
theory of gravity,} J. Phys. A: Math. Gen. 15, 1599 (1982); C.
Barrab\'{e}s and G.F. Bressange, \textit{Singular hypersurfaces in
scalar - tensor theories of gravity,} Class. Quantum Grav. 14, 805
(1997); F. Dahia and C. Romero, \textit{Line sources in
Brans-Dicke theory of gravity}, Phys. Rev. D 60, 104019 (1999).

\bibitem{Sushkov}
S. V. Sushkov, S. M. Kozyrev, \textit{Composite vacuum Brans-Dicke
wormholes}, Phys. Rev. D 84, 124026 .

\bibitem{Vis1}
M. Visser, Phys. Rev. \textit{Traversable wormholes: Some simple
examples}, D 39, 3182–3184 (1989).

\bibitem{Vis2}
M. Visser, \textit{Traversable wormholes from surgically modified
Schwarzschild spacetimes}, Nucl. Phys. B 328, 203-212 (1989).

\bibitem{thin2}
X. Yue, S. Gao, \textit{Stability of Brans-Dicke thin shell
wormholes}, Phys. Lett. A375, 2193-2200 (2011).

\end{thebibliography}
\end{document}